# Функциональные наноструктуры сверхпроводник-ферромагнетик для спинтроники

А. С. Сидоренко[1,2]*, В. Боян[1], Ю.Б.Савва[2], А. Ю. Федотов[3,4], А. В. Вахрушев[3,4,§]

[1]Институт Электронной Инженерии и Нанотехнологий имени Д. Гицу, Молдова, МД-2028, г. Кишинев, ул. Академическая, 3/3
[2]Орловский государственный университет имени И.С. Тургенева, Россия, 302026, г. Орел, ул. Комсомольская, 95
[3]Удмуртский федеральный исследовательский центр Уральского отделения Российской академии наук, Россия, 426067, г. Ижевск, ул. Т. Барамзиной, 34
[4]Ижевский государственный технический университет имени М.Т. Калашникова, Россия, 426069, г. Ижевск, ул. Студенческая, 7

*anatoli.sidorenko@kit.edu , §vakhrushev-a@yandex.ru

Работа посвящена исследованию процессов формирования и анализу параметров функциональной наноструктуры - сверхпроводящего спинового вентиля, представляющего собой многослойную структуру, состоящую из ферромагнитных нанопленок кобальта, разделенных пленками сверхпроводника ниобия. Исследования проводились при помощи моделирования методом молекулярной динамики. Рассмотрены атомарная структура отдельных нанослоев системы. Особое внимание уделено анализу атомарной структуры контактных областей, поскольку качество границы раздела слоев играет решающую роль в создании работоспособного устройства. Реализованы три температурных режима осаждения: 300, 500 и 800 К. Расчеты показали, что при увеличении температуры наблюдается перестроение структуры слоев системы и их разрыхление. Полученные результаты моделирования могут быть использованы как при разработке, так и оптимизации технологий формирования спиновых вентилей и других функциональных элементов спинтроники.

**Введение**

Развитие нанотехнологий привел к возникновению специального раздела квантовой электроники - спинтроники [1]. Основой систем в спинтронике служат гетероструктуры, состоящие из ферромагнетиков, сверхпроводников и нормальных металлов [2], что предполагает создание многослойных нанокомпозитов, образованных нанопленками. Целью настоящего исследования является анализ процессов формирования и структуры сверхпроводящего спинового вентиля на основе многослойной наноструктуры «сверхпроводник (ниобий)-ферромагнетик (кобальт)».

**Математическая модель и постановка задачи**

Изготовление образца происходит вакуумным осаждением материала пленок. В общем случае наносистема содержит более 20 слоев, но процессы их формирования, а также их структурные особенности подобны. Исследование контактного слоя между сверхпроводящими и ферромагнитными материалами осуществлялось методом молекулярной динамики, при этом рассматривался как сам процесс формирования нанослоев, так и полученная результирующая структура, образованная атомами внутри многослойного нанокомпозита. Основу метода молекулярной динамики составляет уравнение движения Ньютона, которое решается для каждой элементарной частицы:

$$m_i \frac{d^2 \mathbf{r}}{dt^2} = -\frac{\partial U(\mathbf{r})}{\partial \mathbf{r}_i} + \mathbf{F}_{ex}, \quad \mathbf{r}_i(t_0) = \mathbf{r}_{i0}, \quad \frac{d\mathbf{r}_i(t_0)}{dt} = \mathbf{V}_{i0}, \quad i = 1, \ldots N \quad (1)$$

где $N$ – общее число атомов наносистемы; $m_i$ – масса $i$-го атома; $\mathbf{r}_{i0}, \mathbf{r}_i(t)$ – начальный и текущий радиус-векторы $i$-го атома соответственно; $U(\mathbf{r})$ – потенциальная энергия или потенциал системы, зависит от взаимного расположения всех частиц; $\mathbf{V}_{i0}, \mathbf{V}_i(t)$ – векторы скорости в начальный и текущий момент; $\mathbf{r}(t) = \{\mathbf{r}_1(t), \mathbf{r}_2(t), \ldots, \mathbf{r}_K(t)\}$ – обобщающая переменная, указывает на зависимость от всех координат атомов; $\mathbf{F}_{ex}$ – сила внешней среды, служит в том числе и для поддержания постоянной температуры. Для определенности решения уравнения молекулярной динамики необходимо наличие уточняющих условий, в качестве которых в (1) выступает указание начальных координат и скоростей для всех атомов.

**Результаты и их анализ**

На рисунке 1 проиллюстрирован послойный анализ гете-роструктуры Nb/Co в вертикальном относительно подложки направлении.

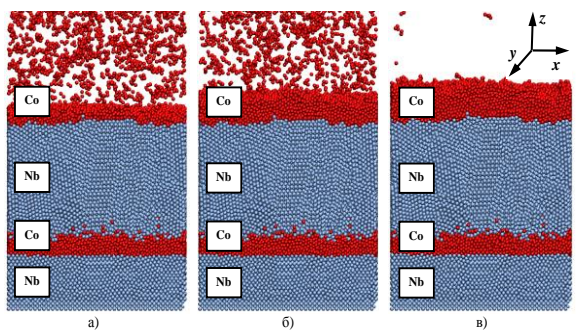

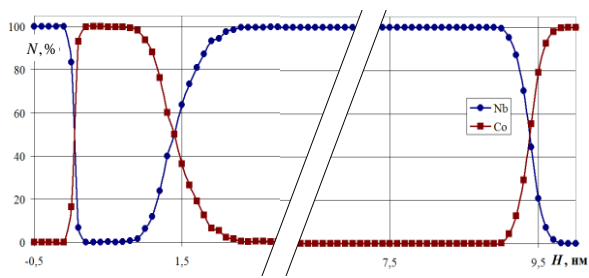

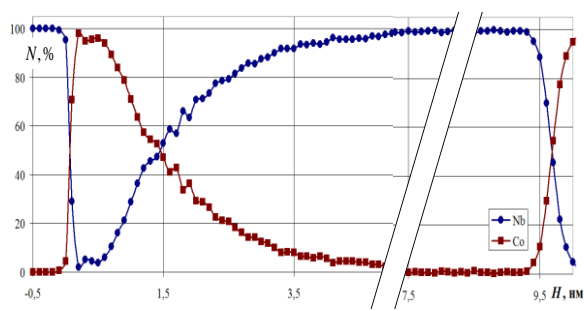

**Рисунок 1.** Осаждение гетероструктуры из ниобия и кобальта при температуре 300 K, время напыления а) 0,1 нс, б) 0,2 нс и в) 0,4 нс ( верхняя панель).
Процентный состав многослойного нанокомпозита Nb/Co, сформированного при температуре 300 K (средняя панель)
Процентный состав многослойного нанокомпозита Nb/Co, сформированного при температуре 800 K (нижняя панель).

В качестве ординаты на графиках указана доля атомов определенного типа, попавшая в тонкий горизонтальный слой толщиной 0,1 нм. Для большей информативности доля атомов отнесена к их общему количеству в текущем слое. Осаждение слоев, соответствующих поочередно чередующимся элементам, Nb и Co , привело к формированию в нанокомпозите контактных слоев - интерфейсов сверхпроводник-ферромагнетик. Области контакта характеризуются сменой структуры и наличием смешанного состава. Первая зона контакта ниобий-кобальт имеет наименьшую размытость, так как формирование первой нанопленки происходит на ровной поверхности подложки. Остальные слои начинают осаждаться на рельефную структуру, образованную на предыдущей стадии моделирования, поэтому рост второй и последующих нанопленок сопровождается протяженными контактными областями. Анализ структуры данных слоев показывает также, что ниобий формируется кристаллическими областями различной направленности. Образование кристаллитов происходит с временным отставанием от поверхностного слоя. Данный эффект связан с последующим перестроением атомов и стремлением их занять энергетически более выгодное состояние. Для нанопленок кобальта характерно строение близкое к аморфному.

В работе были проведены еще две серии вычислительных экспериментов, в которых исследовалось формирование аналогичных гетероструктур при более высоких температурах (500 и 800 K, показанных на рисунке 1 - нижняя панель). Полученные результаты свидетельствуют о существенной зависимости процессов формирования многослойных нанокомпозитов, их контактных областей, а также состава и строения слоев от температуры, при которой происходит формирование многослойной наносистемы.

## Литература